\documentclass[english,superscriptaddress,twocolumn,showpacs,aps,prl]{revtex4-1}
\pdfoutput=1
\usepackage[T1]{fontenc}
\usepackage[latin9]{inputenc}
\usepackage{amsbsy}
\usepackage{amstext}
\usepackage{amssymb}
\usepackage{graphicx}
\usepackage{hyperref}
\hypersetup{
pdftitle={Momentum-Resolved Electronic Structure of the High-Tc Superconductor Parent Compound BaBiO3},
pdfauthor={N. C. Plumb, et al.},
hidelinks,
colorlinks,
allcolors=blue
}
\begin{document}

\title{Momentum-Resolved Electronic Structure of the High-$T_{c}$ Superconductor
Parent Compound BaBiO$_{3}$}

\author{N.~C.~Plumb}
\email{nicholas.plumb@psi.ch}
\affiliation{Swiss Light Source, Paul Scherrer Institut,
CH-5232 Villigen PSI, Switzerland}

\author{D.~J.~Gawryluk}
\altaffiliation{On leave from Institute of Physics, Polish Academy of Sciences, Aleja Lotnikow 32/46, PL-02-668 Warsaw, Poland}
\affiliation{Laboratory for Scientific Developments and Novel Materials, Paul Scherrer Institut, CH-5232 Villigen PSI, Switzerland}

\author{Y.~Wang}
\affiliation{Department of Physics and Astronomy, University of Tennessee, Knoxville,
Tennessee 37996-1200, USA}

\author{Z.~Risti\'{c}}
\author{J.~Park}
\affiliation{Swiss Light Source, Paul Scherrer Institut, CH-5232 Villigen PSI,
Switzerland}

\author{B.~Q.~Lv}
\affiliation{Beijing National Laboratory for Condensed Matter Physics and Institute
of Physics, Chinese Academy of Sciences, Beijing 100190, China}
\affiliation{Swiss Light Source, Paul Scherrer Institut, CH-5232 Villigen PSI,
Switzerland}

\author{Z.~Wang}
\affiliation{Swiss Light Source, Paul Scherrer Institut, CH-5232 Villigen PSI,
Switzerland}
\affiliation{Department of Quantum Matter Physics, 24 Quai Ernest-Ansermet, 1211
Geneva 4, Switzerland}

\author{C.~E.~Matt}
\author{N.~Xu}
\affiliation{Swiss Light Source, Paul Scherrer Institut, CH-5232 Villigen PSI,
Switzerland}

\author{T.~Shang}
\author{K.~Conder}
\affiliation{Laboratory for Scientific Developments and Novel Materials, Paul
Scherrer Institut, CH-5232 Villigen PSI, Switzerland}

\author{J.~Mesot}
\affiliation{Paul Scherrer Institut, CH-5232 Villigen PSI, Switzerland}
\affiliation{Institute of Condensed Matter Physics, \'{E}cole Polytechnique F\'{e}d\'{e}rale
de Lausanne (EPFL), CH-1015 Lausanne, Switzerland}
\affiliation{Laboratory for Solid State Physics, ETH Z\"{u}rich, CH-8093 Z\"{u}rich,
Switzerland}

\author{S.~Johnston}
\affiliation{Department of Physics and Astronomy, University of Tennessee, Knoxville,
Tennessee 37996-1200, USA}

\author{M.~Shi}
\affiliation{Swiss Light Source, Paul Scherrer Institut, CH-5232 Villigen PSI,
Switzerland}

\author{M.~Radovi\'{c}}
\affiliation{Swiss Light Source, Paul Scherrer Institut, CH-5232 Villigen PSI,
Switzerland}
\affiliation{SwissFEL, Paul Scherrer Institut, CH-5232 Villigen PSI, Switzerland}

\date{\today}

\begin{abstract}
We investigate the band structure of BaBiO$_{3}$, an insulating parent
compound of doped high-$T_{c}$ superconductors, using \emph{in situ}
angle-resolved photoemission spectroscopy on thin films. The data
compare favorably overall with density functional theory calculations
within the local density approximation, demonstrating that electron
correlations are weak. The bands exhibit Brillouin zone folding consistent
with known BiO$_{6}$ breathing distortions. Though the distortions are often thought to coincide with  Bi$^{3+}$/Bi$^{5+}$ charge ordering, core level spectra show that bismuth is monovalent. We further demonstrate that the bands closest to the Fermi level are primarily oxygen derived, while the
bismuth $6s$ states mostly contribute to dispersive bands at deeper
binding energy. The results support a model of Bi-O charge transfer
in which hole pairs are localized on combinations of the O $2p$ orbitals.
\end{abstract}


\maketitle

A central challenge for understanding unconventional and/or high-$T_{c}$
superconductors is elucidating how superconductivity emerges from
adjacent phases whose interactions might foster \cite{Anderson1987,Scalapino1995a,Beal-Monod1986,Mazin2009}
or inhibit \cite{Chang2012b,Ghiringhelli2012} electron pairing. Addressing
this issue can be difficult, since often those phases---Mott insulator, spin or charge density wave, ``strange metal,'' and so on---are complex in their
own rights. In this Letter, we perform angle-resolved photoemission
spectroscopy (ARPES) \emph{in situ} on thin films to reveal the electronic
structure of BaBiO$_{3}$, which is an insulating parent compound
of superconductors with $T_{c}$ exceeding 30 K upon doping. Our results
indicate that BaBiO$_{3}$ is weakly correlated and characterized
by a reverse (negative) Bi-O charge transfer, thus supporting a model
in which hole pairs are condensed on combinations of the surrounding
oxygen $2p$ orbitals \cite{Foyevtsova2015}. The long-sought data
are essential information for developing an understanding of the insulating
state, its doping evolution, and superconductivity in bismuthates.

\begin{figure}[b]
\includegraphics[width=0.9\columnwidth]{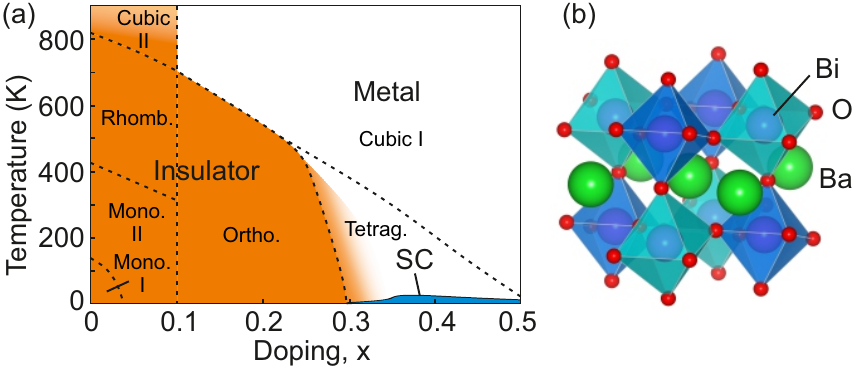}
\caption{\label{fig:phase-diagram}(Color online) (a) Sketch of the $x$-$T$
phase diagram of Ba$_{1-x}$K$_{x}$BiO$_{3}$. The space groups are: Mono.~I
= $P2_{1}/n$; Mono.~II = $I2/m$; Rhomb.~= $R\bar{3}$; Ortho.~=
$Ibmm$; Tetrag.~= $I4/mcm$; Cubic I = $Pm\bar{3}m$; Cubic II = $Fm\bar{3}m$.
(b) Depiction of the insulating ground state ($x=0$) exhibiting breathing
and tilting distortions. The collapsed and expanded BiO$_{6}$ octahedra
are shown in dark and light blue (gray), respectively. In films, the
tilting distortion is suppressed \cite{Inumaru2008}.}
\end{figure}

Figure \ref{fig:phase-diagram}(a) sketches the phase diagram of Ba$_{1-x}$K$_{x}$BiO$_{3}$ \cite{Sleight2015},
which reaches the highest superconducting transition temperature among
the bismuthates ($T_{c}^{\text{max}}=32$ K). Pure BaBiO$_{3}$ is
insulating to well above 800 K \cite{Munakata1992}. This electronic phase extends in an arc in the $T$-vs-$x$ phase diagram out to $x\sim0.3$. The insulating behavior is not only robust in terms of the high metal-insulator transition temperature and persistence to high doping, but also with respect to multiple structural transitions within the insulating dome, which correspond to various degrees of breathing and tilting distortions of the BiO$_{6}$ octahedra. These distortions are depicted in Fig.~\ref{fig:phase-diagram}(b) for the undoped system. The superconducting dome emerges with higher doping, with optimal $T_{c}$ near $x=0.38$. At relatively high doping and low temperature, the cubic metallic phase adopts a tetragonal structure en route to superconductivity. 

The origin and electronic nature of BaBiO$_{3}$'s insulating phase
has been widely discussed. One would naively expect BaBiO$_{3}$ to
be metallic with a half-filled $6s$ band, but the corresponding Bi$^{4+}$
oxidation state is not found in nature. Thus the structure of alternating expanded and collapsed octahedra has often been interpreted as evidence of a corresponding charge-ordered state comprised of Bi$^{3+}$ and Bi$^{5+}$ sites \cite{Cox1976}. However, with exceptions that we will later address \cite{Jeon1990,Kulkarni1990}, experiments have not observed distinct bismuth valences in BaBiO$_{3}$ \cite{Hair1973,Wertheim1982,Shen1990,Boyce1990,Akhtar1993,Namatame1994,Ignatov1996}.
Some theories have modeled the insulating state in terms of an attractive
(negative) effective on-site interaction, $U$ \cite{Rice1981,Varma1988}.
It has also been proposed that holes reside on the oxygen ligands
\cite{Shen1990,Ignatov1996,Menushenkov2000}, implying the influence
of strong Bi-O hybridization or even a negative charge transfer energy.
A recent computational study has advanced the negative charge transfer
view, arguing that the split bands around $E_{F}$ heavily derive
from molecularlike combinations of O $2p$ orbitals while the Bi
$6s$ states mostly contribute to dispersive bands located at deeper
binding energy \cite{Foyevtsova2015}.

Given the questions raised about the role of (negative) effective
on-site interactions in BaBiO$_{3}$, it is fair to ask whether single-electron
models can accurately describe the band structure. Indeed, only recently
have some calculations succeeded to obtain band gaps that match reasonably
well with experiments \cite{Franchini2010}. Even then, the agreement
between the calculated density of states (DOS) and momentum-integrated
photoemission measurements remains underwhelming. An experimental
determination of the band dispersions can provide better input for
theorists and help judge the merits of various models. 

ARPES is a powerful and direct probe of the electronic structure of solids
in $k$ space. Until now photoemission studies of BaBiO$_{3}$ have
lacked momentum resolution \cite{Wertheim1982,Hegde1989,Jeon1990,Shen1990,Itti1991,Namatame1994}.
In order to obtain the ARPES data here, we overcame issues of charging
and sample surface quality by growing smooth, crystalline thin films
of BaBiO$_{3}$ on conducting, grounded bases and performing the experiments
\emph{in situ} on the freshly prepared samples. The measurements were performed at a sample temperature of 18--20 K. We compare the data
with nonrelativistic density functional theory (DFT) calculations
based on the local density approximation (LDA) using the low-temperature bulk space group $P2_{1}/n$. Additional details
regarding the samples, measurements, and calculations can be found
in the Supplemental Material \cite{SupMat}. 

\begin{figure}[t]
\includegraphics[width=0.9\columnwidth]{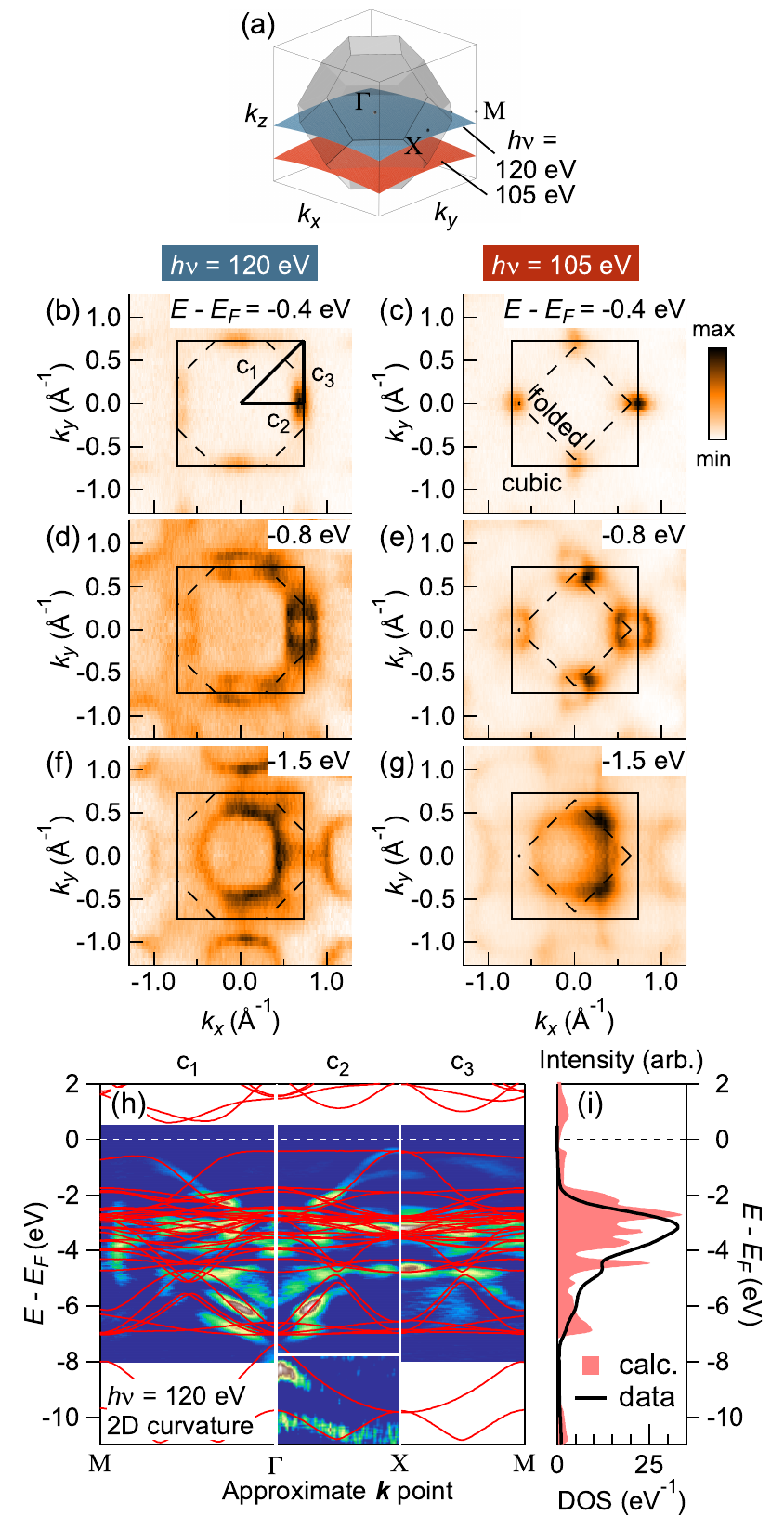}
\caption{\label{fig:electronic-structure}(Color online) (a) Folded Brillouin zone due to breathing distortions of insulating BaBiO$_{3}$. The curved planes show the $k_{z}$ values
accessed by ARPES when using photon energies of 120 eV and 105 eV.
(b)--(g) Left column: ARPES constant energy maps acquired using 120
eV photons, evaluated at $E-E_{F}=-0.4\text{, }-0.8\text{, and}-1.5$
eV; Right column: Analogous maps acquired with $h\nu=105$ eV. (h)
Band dispersions along cuts $c_{1}$, $c_{2}$, and $c_{3}$ indicated
in (b). The data have been analyzed using the 2D curvature method.
LDA calculations of the band structure are overlaid on each cut. (i)
Angle-integrated photoelectron intensity along cut $c_{2}$ (line)
and calculated DOS (shading).}
\end{figure}

The measured electronic structure is summarized in Fig.~\ref{fig:electronic-structure}.
We describe the momentum space in terms of $k_{z}$ perpendicular
to the (001) sample surface and orthogonal $k_{x}$ and $k_{y}$ momenta
lying in the surface plane along the (100) and (010) cubic axes. In
ARPES, a given photon energy, $h\nu$, extracts photoelectrons from
a sheet of $k_{z}$ values scanned in the $k_{x}$-$k_{y}$ plane
\cite{Damascelli2004}. From detailed $h\nu$-dependent scans, we
verified that the observed electronic structure is three dimensional
\cite{SupMat}. As illustrated in Fig.~\ref{fig:electronic-structure}(a),
measurements performed with $h\nu=120$ eV acquire data with $k_{z}$
near the $\Gamma$ point of the 3D Brillouin zone, while $h\nu=105$
eV corresponds to a sheet located close to $k_{z}=-\pi/2a$, where
$a$ is the simple cubic lattice constant ($\approx4.3$ \AA). Figures
\ref{fig:electronic-structure}(b)--\ref{fig:electronic-structure}(g)
show constant energy maps acquired with $h\nu=120$ eV (left column)
and 105 eV (right column) and evaluated at three different binding
energies within the highest occupied bands. In all maps, the counts
were integrated within a range of $\pm15$ meV.

The constant energy maps show that the symmetry of the electronic
structure of the thin film samples is consistent with Brillouin zone folding as depicted in Fig.~\ref{fig:electronic-structure}(a), which arises from oxygen breathing distortions. The different $k_{z}$ values associated
with the two photon energies lead to distinct appearances of the zone folding when viewed in the $k_{x}$-$k_{y}$ plane. Namely, the in-plane projection of the folded Brillouin zone goes from being octagonal for $h\nu=120$ eV to square in the case of $h\nu=105$ eV. These projections are overlaid on their respective maps (dashed lines),
along with the simple cubic Brillouin zone (solid lines).

Analysis of the band dispersions is presented in Fig.~\ref{fig:electronic-structure}(h).
Labels $c_{1}$--$c_{3}$ refer to cuts through the data as marked
in Fig.~\ref{fig:electronic-structure}(b)\emph{.} To clarify the
bands, the spectra have been processed using the 2D curvature method
\cite{Zhang2011b}. The $\boldsymbol{k}$-integrated intensity obtained
from cut $c_{2}$ is shown in Fig.~\ref{fig:electronic-structure}(i).
LDA calculations of the bands and DOS are overlaid on the experimental
data in Figs.~\ref{fig:electronic-structure}(h) and \ref{fig:electronic-structure}(i),
respectively. For the calculations, we eliminated the octahedral tilting
distortion in accordance with the structure of BaBiO$_{3}$ thin films
grown on various substrates \cite{Inumaru2008}. This slightly increases
the bandwidths and alters some band splittings, modestly but noticeably
improving the agreement between the data and LDA. One should note
that the \emph{breathing} distortions still present in films are found
to be the deciding structural factor in BaBiO$_{3}$'s insulating
behavior \cite{Inumaru2008,Thonhauser2006,Foyevtsova2015}. Hence
the general conclusions of the present study are believed to be applicable
to bulk BaBiO$_{3}$.

The level of agreement between the data and calculations demonstrates
that LDA methods can successfully compute most aspects of the band
structure of BaBiO$_{3}$, including the bandwidths. 
Consequently,
we can infer that electron correlations are weak. 
Certain portions of the calculated electronic structure, such as the flat band crossing $\Gamma$ near $E_F$, are difficult to observe in the experiments. These features tend to be associated with a secondary folding, in addition to that of the breathing distortions, occurring in the low-temperature ``Mono.~I'' $P2_1/n$ space group [Fig.~\ref{fig:phase-diagram}(a)]. Ostensibly the symmetry-breaking potential associated with this secondary folding is weak, leading to the very low spectral weight of these features in ARPES \cite{SupMat}.
Aside from that, the disagreements are quantitative and relatively minor, 
although the LDA tends to underestimate the band gap. Here
the calculated direct gap is 1.45 eV, compared to an experimental
value of $\sim2$ eV in both bulk and thin film samples \cite{Tajima1985,Sato1989}.
Moreover, the calculation finds the indirect gap on the verge of closing,
whereas our experiments show this is not the case \cite{SupMat}.
More costly hybrid functionals might be able to obtain better quantitative
results regarding the band gap \cite{Franchini2010}.

Photoemission spectroscopy from core levels is in principle well suited
to address the issue of formal Bi charge ordering in BaBiO$_{3}$,
but previous studies did not arrive at a consensus. Several measurements
found little evidence of distinct Bi valence states \cite{Wertheim1982,Hegde1989,Shen1990,Namatame1994};
however, the outcomes were sensitive to the sample surface preparations,
and in some cases additional peaks or shoulders were observed \cite{Jeon1990,Kulkarni1990,Nagoshi1992}.
Because of the very short probing depth of photoemission techniques, it
has not been clear which of these measurements---if any---represent
the intrinsic bulklike electronic structure rather than some extrinsic
and/or surface-related system that was perhaps contaminated, damaged, or structurally distorted. 

\begin{figure}[b]
\includegraphics[width=0.9\columnwidth]{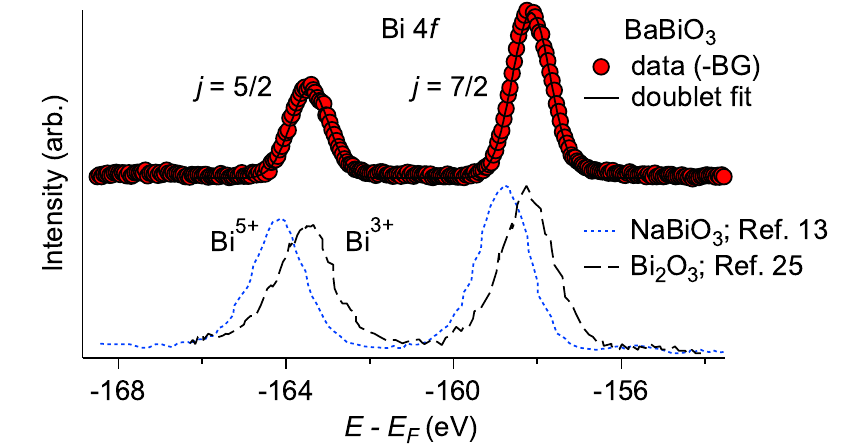}
\caption{\label{fig:xps}(Color online) Bi $4f$ core level spectrum.
Following background subtraction ($-$BG), the data (circles) are fit
by two Gaussians (line). Spectra from Bi$_{2}$O$_{3}$ (dashed line;
Ref.~\cite{Hegde1989}) and NaBiO$_{3}$ (dotted line; Ref.~\cite{Kulkarni1990}),
representing cases of pure Bi$^{3+}$ and Bi$^{5+}$ valence states,
respectively, are shown for comparison.}
\end{figure}

Having demonstrated the 3D insulating band structure of the BaBiO$_{3}$
films, which matches well with calculations, the present experiments
allow us to definitively assess the Bi core level spectra relevant
to the bulklike electronic structure. Figure \ref{fig:xps} shows
the Bi $4f$ spectrum acquired with $h\nu=220$ eV. No evidence of
two distinct Bi charge states is seen; rather, the $j=5/2$, $7/2$
doublet is well fit by a double-Gaussian function. The peaks from
BaBiO$_{3}$ are sharp in comparison to published data from Bi$_{2}$O$_{3}$
\cite{Hegde1989} and NaBiO$_{3}$ \cite{Kulkarni1990} (nominal representatives
of pure Bi$^{3+}$ and Bi$^{5+}$ valence states, respectively), which
argues against mixed or fluctuating valence of the Bi atoms.

\begin{figure*}[t]
\includegraphics[width=\textwidth]{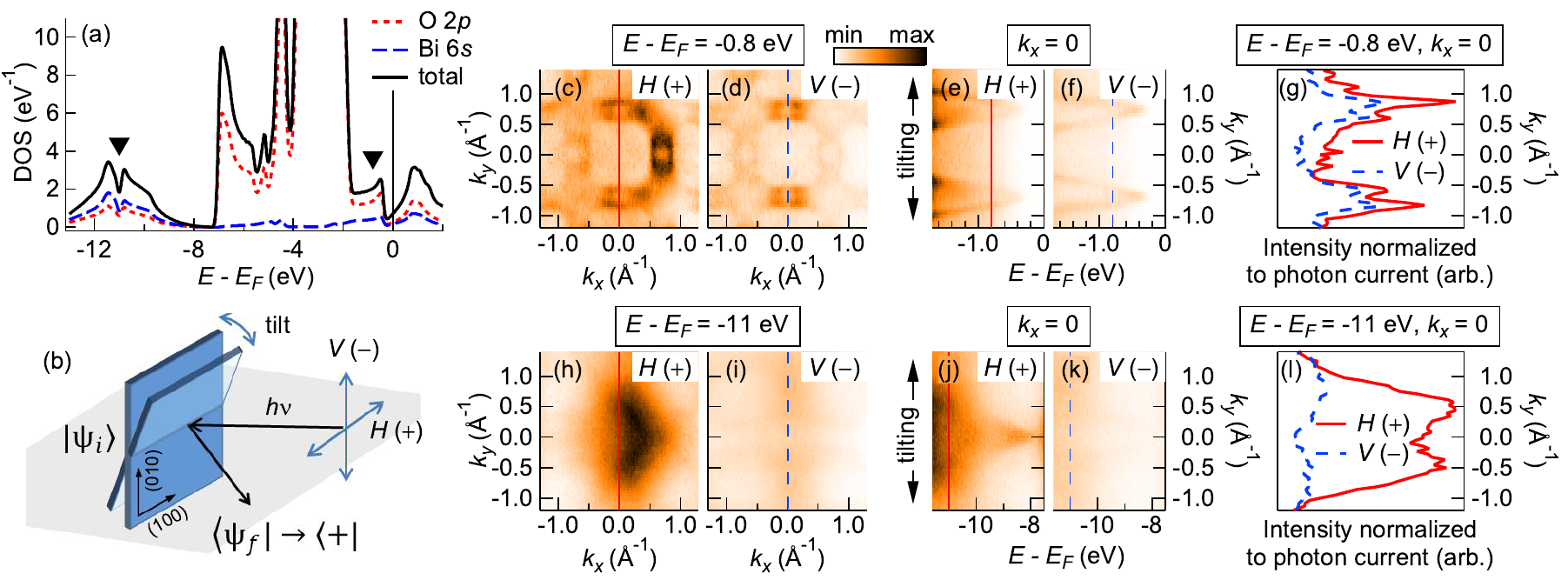}
\caption{\label{fig:orbital-symmetry}(Color online) (a) Calculated total DOS (solid line), as well as the O $2p$ and Bi $6s$
orbital-projected DOS (short and long dashed lines, respectively).
(b) Sketch of the experimental geometry. (c), (d) Constant energy
maps at $E-E_{F}=-0.8$ eV acquired using $h\nu=120$ eV with $H$ and
$V$ photon polarizations. (e), (f) Near-$E_{F}$ dispersion cuts through
$k_{x}=0$ acquired with $H$ and $V$ polarizations. The momentum $k_{y}$
was varied by tilting the sample. (g) MDCs at $k_{x}=0$ and $E-E_{F}=-0.8$
eV obtained with $H$ and $V$ polarizations (solid and dashed lines, respectively).
The intensities have been normalized to the incident photon current
of each polarization and shifted by an overall background offset.
(h)--(l) Analogous to (c)--(g), except analyzed at deeper binding
energies, as indicated.}
\end{figure*}

To further uncover the electronic nature of BaBiO$_{3}$ and understand
why bismuth charge ordering is not observed, we turn attention to
the orbital compositions of the bands. Figure \ref{fig:orbital-symmetry}(a) shows the total and orbital-projected DOS computed from the band structure in Fig.~\ref{fig:electronic-structure}(h). The calculations find that the highest occupied bands derive overwhelmingly from O $2p$ states, while most of Bi $6s$ states contribute to dispersive bands roughly 8 to 13 eV below $E_{F}$, accounting for the majority of the DOS in that energy range. These deep-energy bands, which are nearly invisible in momentum-integrated photoemission, can be unambiguously identified in the ARPES data {[}compare Figs.~\ref{fig:electronic-structure}(h) and \ref{fig:electronic-structure}(i){]}. This allows us to probe the photoemission matrix elements of these bands in order to check consistency with the calculated orbital hierarchy. 

In our experimental geometry {[}Fig.~\ref{fig:orbital-symmetry}(b){]},
the photoexcitation operator, $\boldsymbol{A}\cdot\boldsymbol{p}$,
has even parity ($+$) with respect to the beam-detector plane when
the radiation is horizontally ($H$) polarized and odd parity ($-$)
for vertical ($V$) polarization. Regarding the detected photoelectron
final state as a plane wave \cite{Damascelli2004}
, the use of V polarized photons will cause the matrix element $M_{fi}\propto\left|\left\langle \psi_{f}\right|\boldsymbol{A}\cdot\boldsymbol{p}\left|\psi_{i}\right\rangle \right|^{2}$
to vanish when the initial state is even (i.e., $M_{fi}\rightarrow\left|\left\langle +\right|-\left|+\right\rangle \right|^{2}=0$).
By contrast, the same state would generally be visible when probed
in the $H$ polarization mode. Thus, as a telling signature, a majority
$s$-orbital character band should exhibit strong tilt-independent
suppression of the photoemission intensity upon switching from $H$ to
$V$ polarization. A band composed of combinations of $p$ orbitals,
on the other hand, would display more complicated tilt-dependent intensity differences between the two polarization modes, since the orbitals' parities depend on the geometry. 

As shown in Figs.~\ref{fig:orbital-symmetry}(c) and \ref{fig:orbital-symmetry}(d), constant energy maps at $E-E_{F}=-0.8$ eV acquired with $H$ polarization are similar to those obtained using $V$ polarization, simply differing in their intensity modulations as a function of the geometry. In dispersion cuts at $k_{x}=0$, which are obtained by tilting the sample, the highest occupied band is clearly visible over its full bandwidth for both incoming polarizations {[}Figs.~\ref{fig:orbital-symmetry}(e) and \ref{fig:orbital-symmetry}(f){]}. The peak features seen in momentum distribution curves (MDCs) evaluated at $E-E_{F}=-0.8$ eV and $k_{x}=0$, shown in Fig.~\ref{fig:orbital-symmetry}(g), have similar intensities for both polarizations. The behavior is markedly different at deeper
binding energies of $E-E_{F}\lesssim-8$ eV, where the features seen
in the $H$ mode nearly vanish when probed with $V$ polarization {[}Figs.~\ref{fig:orbital-symmetry}(h)--\ref{fig:orbital-symmetry}(l){]}.
The strong tilt-independent suppression of spectral weight observed
from the bands at deep binding energy is consistent with a majority
$s$-orbital character and serves as further validation of the LDA-computed DOS.

The determined orbital hierarchy, in which O $2p$ states dominate
at the gap edge and the Bi $6s$ states are concentrated far below
$E_{F}$, indicates that holes in BaBiO$_{3}$ reside primarily on
the oxygen orbitals. This, in turn, provides an explanation for the
lack of bismuth charge ordering: hole pairs from the would-be Bi$^{5+}$
sites of the collapsed octahedra are transferred to the oxygen ligands,
leaving behind just Bi$^{3+}$, consistent with core level measurements
(Fig.~\ref{fig:xps}). LDA calculations of closely related SrBiO$_{3}$
found that these holes specifically occupy $A_{1g}$ combinations
of the O $2p$ orbitals within the sublattice of collapsed BiO$_{6}$
octahedra \cite{Foyevtsova2015}. We note that the resulting electronic
configuration resembles an $s$-$p$ analog of a model
proposed for nickelates \cite{Mizokawa2000,Park2012a,Lau2013,Johnston2014}.

The overall agreement between the measured and calculated band structure
builds confidence in conclusions drawn from DFT studies. One clear
implication is that strong nesting of the underlying Fermi surface
of cubic BaBiO$_{3}$ leads to a large static susceptibility peaked
at $Q=(\pi,\pi,\pi)/a$ \cite{Mattheiss1983,Sahrakorpi2000,Foyevtsova2015}.
The charge density wave state in BaBiO$_{3}$ is remarkable,
however, in terms of its fully gapped 3D band structure and its temperature
and doping stability. These properties might derive in part from influences
beyond a pure Peierls model. Specifically, added holes have been proposed
to form trapped (bi)polarons inside the gap \cite{Bischofs2002},
which is likewise supported by the application of DFT \cite{Franchini2009}.

These and other studies \cite{Zhao2000} certainly point toward strong
electron-phonon interactions in BaBiO$_{3}$, although it is still
not clear whether the high optimal $T_{c}$ in, e.g., Ba$_{1-x}$K$_{x}$BiO$_{3}$
can be entirely explained within BCS strong coupling theory \cite{Bazhirov2013,Yin2013}.
For example, as (bi)polarons become mobile with increased hole doping,
superconductivity in the bismuthates could involve \emph{local} bosonic
pairs \cite{Micnas1990,Mott1993}. In many cases, bipolaronic superconductivity
has been discussed in the context of a classically charge-ordered
state in BaBiO$_{3}$, regarding electron and hole pairs as localized
on the supposed Bi$^{3+}$ and Bi$^{5+}$ sites. Our data demonstrate
that in the undoped ground state \emph{all} Bi atoms have essentially
filled $6s$ shells, and the Bi sites of the collapsed and expanded
octahedral sublattices differ in terms of their screening by holes
distributed on the surrounding oxygen orbitals. Theories of superconductivity
in bismuthates (bipolaronic or otherwise) may benefit from taking
this more accurate description of the electronic configuration and
screening into account.

In conclusion, by performing \emph{in situ} spectroscopy on freshly
grown films, we have obtained the first ARPES measurements from the
high-$T_{c}$ parent compound BaBiO$_{3}$. The data are in good agreement
with DFT-LDA calculations. The simultaneous observations of the folded
Brillouin zone due to oxygen breathing distortions and sharp, single-component
Bi core levels demonstrate that the structurally inequivalent Bi sites
have virtually identical valences. In the calculations, the bands
closest to $E_{F}$ are primarily derived from O $2p$ states, while
the Bi $6s$ states mostly contribute to the deep-energy band structure,
which is shown to be consistent with the behavior of the ARPES matrix
elements. The results signal the influence of a negative charge transfer
energy, driving hole pairs to occupy ligand states on the collapsed
octahedra. The findings should be relevant for understanding the doping
evolution and superconducting behavior of doped bismuthate compounds.

\bibliographystyle{apsrev4-1}
\bibliography{citations}

\begin{acknowledgments}
The authors thank G.~A.~Sawatzky, K.~Foyevtsova, and C.~Mudry
for insightful discussions. F.~Dubi, M.~Kropf, and L.~Nue provided
technical assistance at SIS beam line. D.~J.~G.~received financial
support from Sciex-NMSch (Project No.~13.236) funded by the Swiss
Confederation. Y.~W.~and S.~J.~were supported by the University of Tennessee's 
Science Alliance Joint Directed Research and Development (JDRD) program, 
a collaboration with Oak Ridge National Laboratory. S.~J.~acknowledges 
additional support from the University of Tennessee's Office of Research \& 
Engagement's Organized Research Unit program. CPU time was provided in part by resources supported by the University of Tennessee and Oak Ridge National Laboratory Joint
Institute for Computational Sciences (\href{http://www.jics.utk.edu}{http://www.jics.utk.edu}).
\end{acknowledgments}

\end{document}